\documentclass{ws-procs9x6}

\begin{document}

\title{Lattice QCD Study of the Pentaquark Baryons}

\author{T.~T.~Takahashi, T.~Umeda, T.~Onogi and T.~Kunihiro}

\address{Yukawa Institute for Theoretical Physics, Kyoto University,
Kitashirakawa-Oiwakecho, Sakyo, Kyoto 606-8502, Japan}

\maketitle

\abstracts{
We study the spin $\frac12$ hadronic state in
quenched lattice QCD to search for a possible $S=+1$
pentaquark resonance.
Simulations are carried out
on $8^3\times 24$, $10^3\times 24$, $12^3\times 24$
and $16^3\times 24$ lattices at $\beta$=5.7 at the quenched level
with the standard plaquette gauge action and Wilson quark action.
We adopt two independent operators with $I=0$ and $J^P=\frac12$
to construct a $2\times 2$ correlation matrix.
After the diagonalization of the correlation matrix,
we successfully obtain the energies of the ground-state and 
the 1st excited-state in this channel.
The volume dependence of the energies 
suggests the existence of 
a possible resonance state slightly above
the NK threshold in $I=0$ and $J^P=\frac12^-$ channel.
}

\section{Introduction}\label{Intro}
After the discovery~\cite{Netal03} of $\Theta^+(1540)$,
the property of the particle is one of the central issues
in hadron physics.
While it is very likely that the isospin of $\Theta^+$ is zero,
the spin and the parity of $\Theta^+$ and the origin
of its tiny width of a few MeV still remain open questions.~\cite{O04}
In spite of many theoretical studies on $\Theta^+$,~\cite{O04}
the nature of the exotic particle is still controversial.

As for the theoretical approaches,
the lattice QCD is considered as one of the
most reliable tools for studying the properties of hadron states
from first principle, which has been very successful
in reproducing the non-exotic hadron mass spectra.~\cite{CPPACS01}
Up to now, several lattice QCD studies have been reported,
which mainly aim to look for pentaquarks.
However, the results are unfortunately contradictory with each other.
On one hand, in refs.~\cite{CFKK03,S03}, the authors conclude
the parity of $\Theta^+$ is negative.
On the other hand, in ref.~\cite{CH04},
$\Theta^+$ in the positive parity channel is reported.
In refs.~\cite{Metal04,IDIOOS04}, even the absence of $\Theta^+$ is suggested.

A difficulty in the study of $\Theta^+$ using lattice QCD
comes from the contaminations of the scattering states of nucleon and Kaon,
since the mass of $\Theta^+$ is above the NK threshold.
In order to verify the existence of a resonance state,
we need to properly discriminate the first few lowest states,
pick up one state as a candidate, and 
determine whether it is a resonance state or not.

In this paper, 
we study $I=0$ and $J=\frac12$ channel in quenched lattice QCD.
With the aim to extract the possible resonance state 
slightly above the NK threshold in this channel,
we adopt two independent operators with $I=0$ and $J=\frac12$
and diagonalize the $2\times 2$ correlation matrix of the operators.
After the successful separation of the states, 
we investigate the volume dependence of the energy
so that we can distinguish the resonance state 
from simple scattering states.

\section{Formalism to separate the states}\label{Formalism}

In lattice QCD calculations,
the energies of states can be measured using the operator correlations.
Let us consider the correlation
$\langle O(T)O(0)^\dagger\rangle$ with $O(T)$
an interpolating operator 
on the time plane $t=T$ with a certain quantum number.
The correlation $\langle O(T)O(0)^\dagger\rangle$ physically represents
the situation in which
the states with the same quantum number as $O(t)$
are created at $t=0$ and propagate into Euclidean time direction and finally
are annihilated at $t=T$.
Taking into account the fact that the created state $O^\dagger|vac\rangle$
can be expressed by the complete set of QCD eigenstates as
$O^\dagger|vac\rangle = c_0|0\rangle +c_1|1\rangle +.. =\sum c_i|i\rangle$,
we can write down the correlation $\langle O(T)O(0)^\dagger\rangle$
as $\langle O(T)O(0)^\dagger\rangle = 
\sum_i \sum_j \bar{c}_ic_j \langle i|e^{-\hat HT}|j\rangle
= \sum_i |c_i|^2e^{-E_iT}$ in terms of the QCD transfer matrix $e^{-\hat HT}$,
with QCD Hamiltonian $\hat H$
and $n$-th excited-state $|n\rangle$
normalized as $\langle m|n\rangle =\delta_{mn}$.
We can extract the ground-state energy $E_0$
by taking large $T$ so that
the correlation $\langle O(T)O(0)^\dagger\rangle$ behaves
as a single-exponential function
$\langle O(T)O(0)^\dagger\rangle\sim |c_0|^2e^{-E_0T}$.
It is however rather difficult to obtain the excited-state energy $E_i$($i>0$)
from the multi-exponential function of $T$,
which is a problem for extracting the possible pentaquark signal
above the threshold.
To overcome this difficulty, 
we adopt the variational method using correlation matrices
of independent operators.~\cite{TS04}
In this method, we need a set of independent operators \{$O^I$\}.
We define the correlation matrix ${\bf C}_{IJ}$ as
${\bf C}_{IJ}(T)\equiv
\langle O^I(T) O^{J\dagger}(0)\rangle$.
Then, from the product ${\bf C}^{-1}(T+1){\bf C}(T)$,
we can extract the energies \{$E_i$\} as
the logarithm of eigenvalues \{$e^{E_i}$\} of the matrix 
${\bf C}^{-1}(T+1){\bf C}(T)$.
In the case when we prepare N independent operators,
the correlation matrix is an $N\times N$ matrix and
we can obtain N eigenenergies \{$E_i$\} ($0\leq i\leq N-1$).

\section{Lattice set up}\label{Setup}

As interpolating operators $\{O^I\}$,
we take two independent operators;
$\Theta^1(x)\equiv 
\varepsilon^{abc}[u_a^{\rm T}(x)C\gamma_5d_b(x)]u_e(x)
[{\overline{s_c}(x)\gamma_5d_e(x)}]-(u\leftrightarrow d)$
which is expected to have a larger overlap with $\Theta^+$ state,
and $\Theta^2(x)\equiv 
\varepsilon^{abc}
[u_a^{\rm T}(x)C\gamma_5d_b(x)]u_c(x)
[{\overline{s_e}(x)\gamma_5d_e(x)}]-(u\leftrightarrow d)$ 
which we expect to have larger overlaps with NK scattering states.
Both of them have the quantum number of $(I,J)=(0,\frac12)$.
Here, Dirac fields 
$u(x)$, $d(x)$ and $s(x)$ are up, down and 
strange quark field, respectively and 
the Roman alphabets
\{a,b,c,e\} denote color indices.
In this case, the correlation matrix is a $2\times 2$ matrix.
Furthermore, for a source, we construct and use ``wall'' operators
$\Theta^1_{\rm wall}(t)$ and 
$\Theta^2_{\rm wall}(t)$
defined using spatially spread quark fields
$\sum_{\vec{\bf x}}q(x)$ as
$\Theta^1_{\rm wall}(t)\equiv \left(\sqrt{\frac{1}{V}}\right)^5
\sum_{\vec{\bf x_1}\sim \vec{\bf x_5}}
\varepsilon^{abc}
[u_a^{\rm T}(x_1)C\gamma_5
d_b(x_2)]u_e(x_3)[{\overline{s_c}(x_4)
\gamma_5 d_e(x_5)}]-(u\leftrightarrow d)$.
We fix the source operator $\overline{\Theta}_{\rm wall}(t)$
on t=6 plane to reduce the effect of the Dirichlet boundary
on t=0 plane,~\cite{CPPACS04}
and calculate 
${\bf C}^{IJ}(T)=
\sum_{\vec{\bf x}} \langle
\Theta^I(\vec{\bf x},T+6) \overline{\Theta}^{J}_{\rm wall}(6)\rangle$.

We adopt the standard plaquette (Wilson) gauge action at $\beta=5.7$
and Wilson quark action with the hopping parameters as
$(\kappa_{u,d},\kappa_{s})$=$(0.1600,0.1650)$, $(0.1625,0.1650)$,
$(0.1650,0.1650)$, $(0.1600,0.1600)$ and $(0.1650, 0.1600)$,
which correspond to the current quark masses 
$(m_{u,d},m_{s})\sim(240,100)$, $(170,100)$,
$(100,100)$, $(240,240)$ and $(100,240)$, respectively in the unit of MeV.
We perform lattice QCD calculations at $\beta=5.7$ 
employing four different sizes of lattices,
$8^3\times 24$, $10^3\times 24$, $12^3\times 24$ and $16^3\times 24$
with 2900, 2900, 1950 and 950 gauge configurations.
At $\beta=5.7$, the lattice spacing $a$
is set to be 0.17 fm so that the $\rho$ meson mass is 
properly reproduced.
Then, in the physical unit, the lattice sizes are
$1.4^3\times 4.0$ fm$^4$, $1.7^3\times 4.0$ fm$^4$, 
$2.0^3\times 4.0$ fm$^4$ and $2.7^3\times 4.0$ fm$^4$.

We take periodic boundary conditions in the directions for the gauge field,
whereas
we impose periodic boundary conditions on the spatial directions
and the Dirichlet boundary condition on the temporal direction 
for the quark field.
Many lattice studies about $\Theta^+$
adopted a periodic or antiperiodic boundary condition
on the temporal direction for quarks. 
In the case when the temporal length $N_t$ is not long enough,
we have to be careful about the contaminations by particles which
go beyond the temporal boundary.
For example, with the periodic boundary condition,
the correlation $\langle \Theta(T)\overline{\Theta} (0)\rangle$ contains 
the unwanted contributions such as
$\langle K|\Theta(T)|N\rangle \langle N|\overline{\Theta} (0)|K\rangle
\sim e^{-E_NT+E_K(T-N_t)}$.
While no quark can go over the boundary on t=0 in the temporal direction,
the boundary is transparent 
for the particles composed only by gluons; {\it i.e.} glueballs,
due to the periodicity of the gauge action.
Taking into account that these particles are rather heavy,
it would be however safe 
to neglect these gluonic particles going beyond the boundary.
Then, the correlation $\langle \Theta(T)\overline{\Theta} (0)\rangle$
mainly contains such terms as $\langle vac|\Theta(T)|NK\rangle \langle
NK|\overline{\Theta} (0)|vac\rangle$ and
we can use the exactly same prescription mentioned in the previous section.

\section{Ground-state and 1st Excited-state in $(I,J^P)=(0,\frac12^-)$
 channel}
\label{Negative}

In this section, we investigate 
the ground-state and the 1st excited-state in $(I,J^P)=(0,\frac12^-)$ channel.
We focus on the volume dependence of the energy 
of each state, to distinguish a possible resonance state from NK scattering
states.
We expect a resonance state to have almost no volume dependence,
while the energies of NK scattering states are expected to scale as 
$\sqrt{M_N^2+|\frac{2\pi}{L}\vec{\bf n}|^2}+
\sqrt{M_K^2+|\frac{2\pi}{L}\vec{\bf n}|^2}$ according to
the relative momentum $\frac{2\pi}{L}\vec{\bf n}$ between N and K on
finite periodic lattices on the assumption that the
NK interaction is negligible.
Note here that there are other possible finite volume effects 
due to the finite size of N and K or possible pentaquark states,
than the volume dependence arising from $\frac{2\pi}{L}\vec{\bf n}$.
We then need to take account of this fact in the following analysis.

Now we show the lattice QCD results
of the ground-state and the 1st excited-state
in $I=0$ and $J^P=\frac12^-$ channel.
\begin{figure}[htb]
\begin{center}
\includegraphics[scale=0.29]{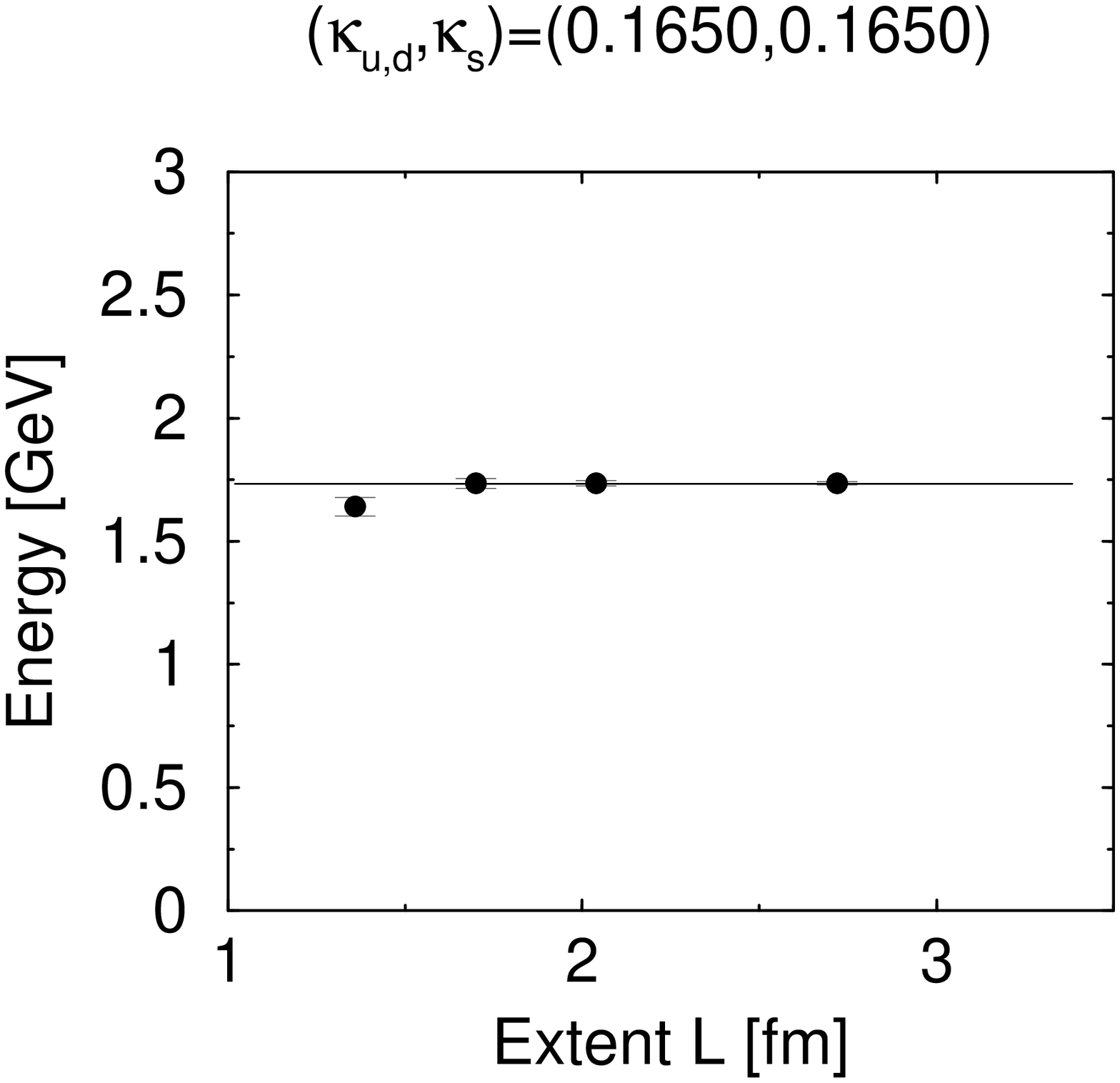}
\includegraphics[scale=0.29]{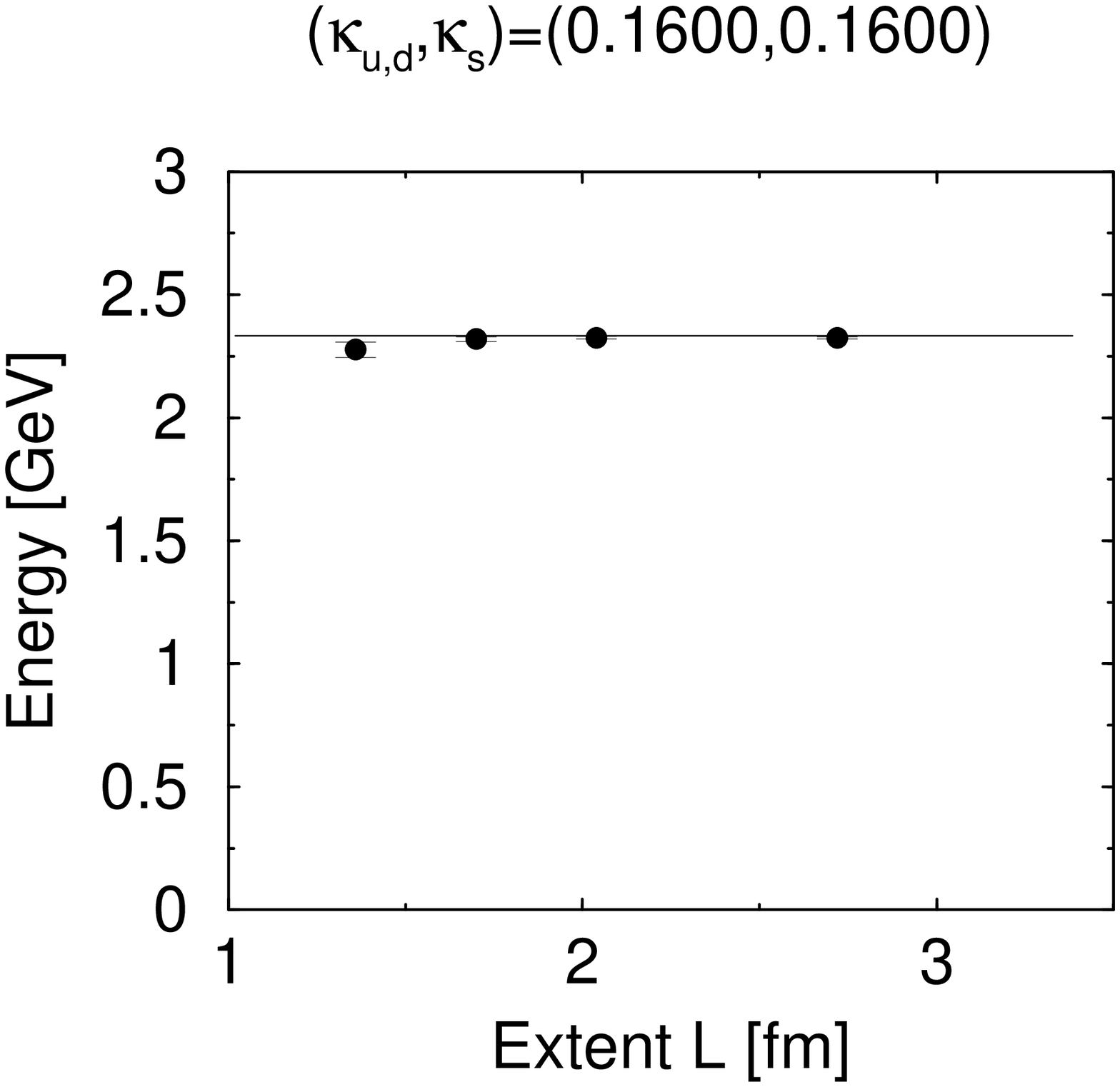}
\end{center}
\caption{\label{negativeGS}
The lattice QCD data of the ground-state in
$(I,J^P)=(0,\frac12^-)$ channel
are plotted against the lattice extent $L$.
The left figure is for $(\kappa_{u,d},\kappa_s)=(0.1650,0.1650)$
and the right figure
is for $(\kappa_{u,d},\kappa_s)=(0.1600,0.1600)$.
The solid line denotes the simple sum $M_N+M_K$
of the masses of nucleon $M_N$ and Kaon $M_K$.
We use the central values of $M_N$ and $M_K$ 
obtained on the lattice with $L=16$
}
\end{figure}
In Fig.~\ref{negativeGS}, we plot the ground-state energies
in $I=0$ and $J^P=\frac12^-$ channel on four different volumes.
Here the horizontal axis denotes the lattice extent $L$
in the physical unit and the vertical axis is the energy of this state.
The symbols with error-bars are lattice data and the solid line
denotes the sum $M_N+M_K$
of the nucleon mass $M_N$ and Kaon mass $M_K$.
At a glance, 
we find that the energy of this state takes almost constant value against
the volume variation and coincides with
the simple sum $M_N+M_K$.
We can then conclude the ground-state in $I=0$ and $J^P=\frac12^-$ channel
is the NK scattering state with the relative momentum $|{\bf p}|=0$.

\begin{figure}[htb]
\begin{center}
\includegraphics[scale=0.29]{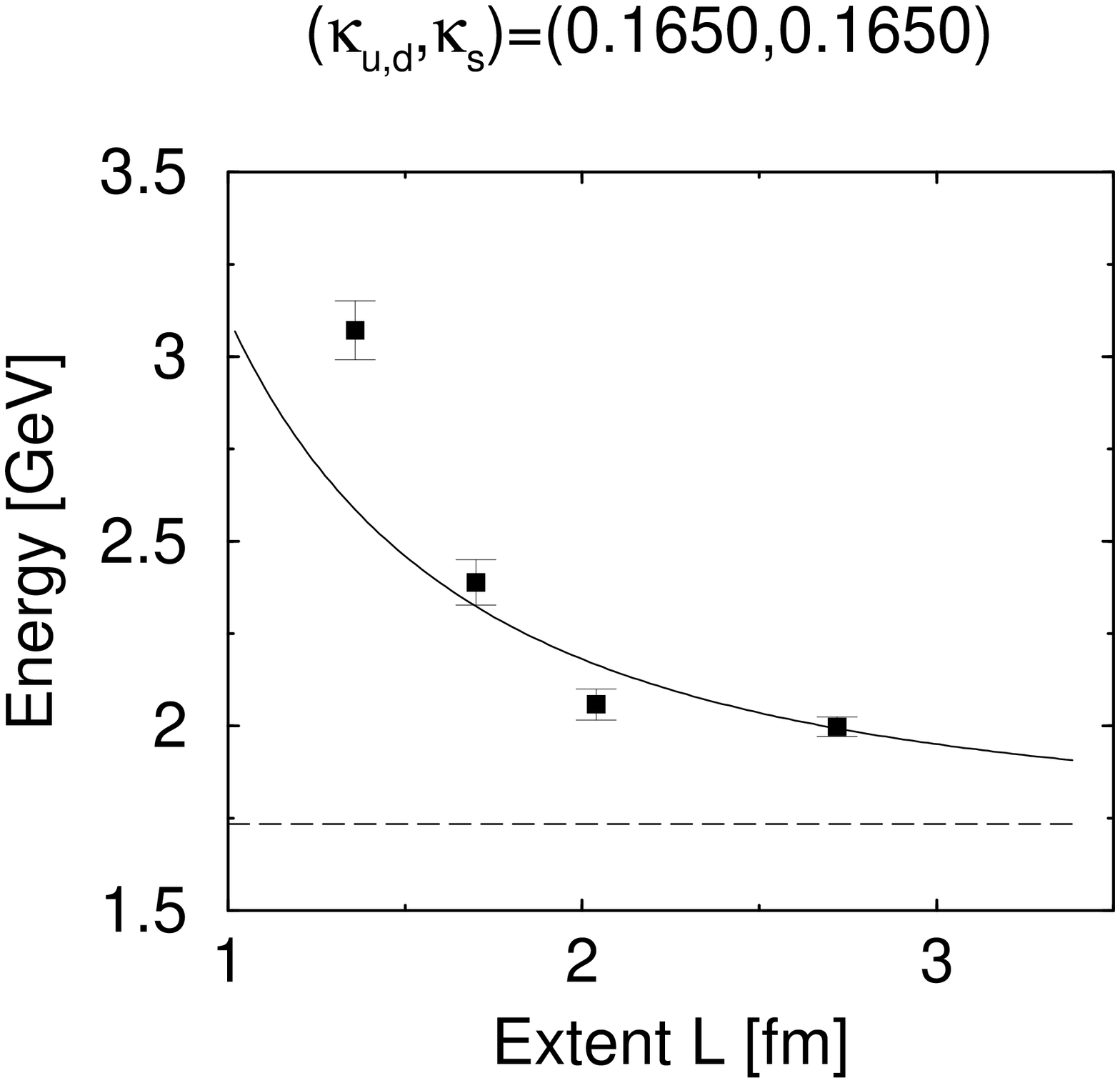}
\includegraphics[scale=0.29]{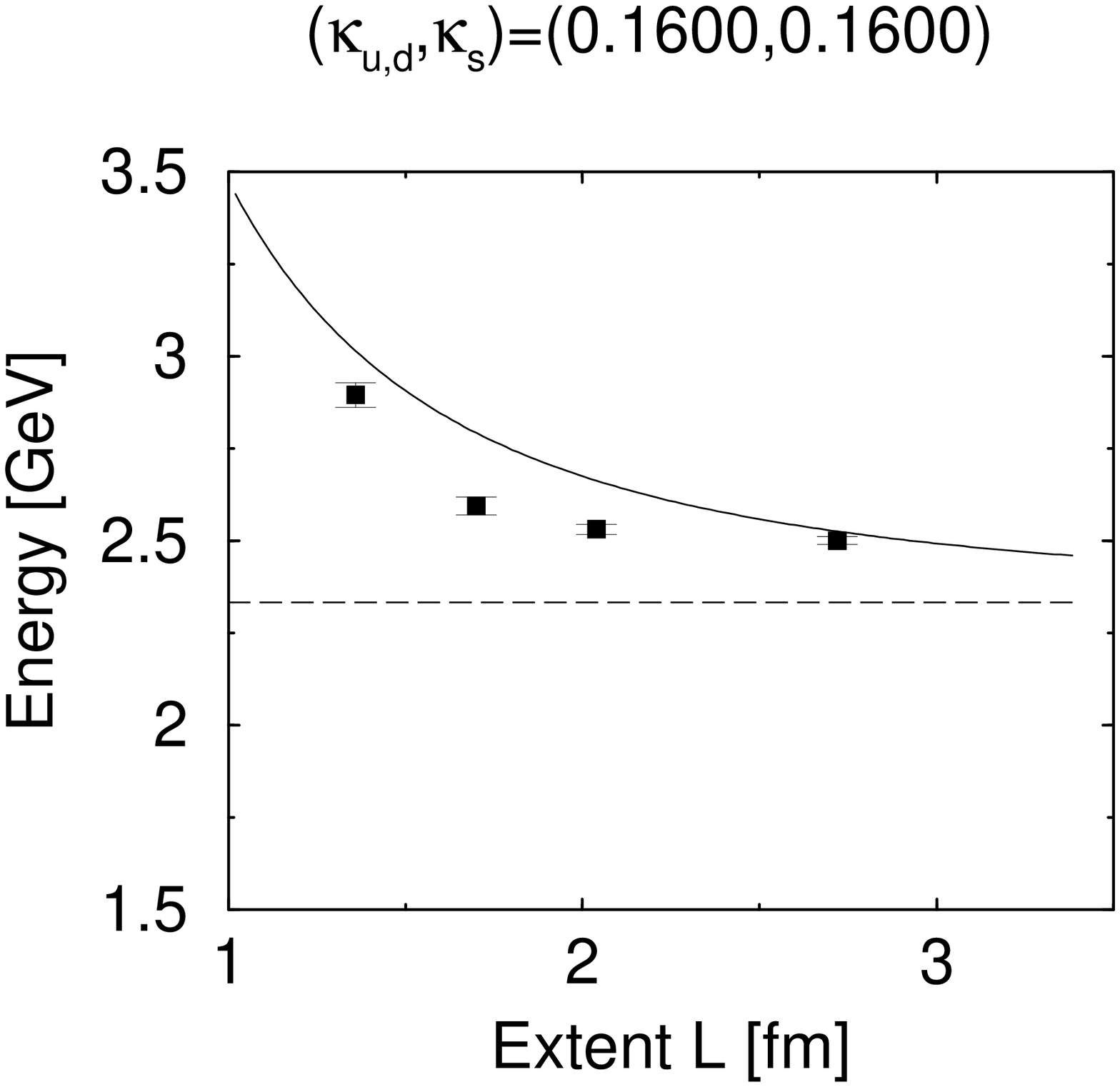}
\end{center}
\caption{\label{negativeES}
The lattice QCD data of the 1st excited-state in
$(I,J^P)=(0,\frac12^-)$ channel
are plotted against the lattice extent $L$.
The solid line represents
$\sqrt{M_N^2+|{\bf p}|^2}+\sqrt{M_K^2+|{\bf p}|^2}$
with $|{\bf p}|=2\pi/L$ the smallest relative momentum
on the lattice.
The dashed line denotes $M_N+M_K$.
}
\end{figure}
Fig.~\ref{negativeES}
shows the 1st excited-state energies in this channel
again in terms of $L$.
The solid line shows $\sqrt{M_N^2+(2\pi/L)^2}+\sqrt{M_K^2+(2\pi/L)^2}$
as the function of $L$. 
In this case, the lattice data exhibit clearly some volume dependence.
One would naively consider this dependence to be that of
the 2nd lowest NK scattering state,
which scales as $\sqrt{M_N^2+(2\pi/L)^2}+\sqrt{M_K^2+(2\pi/L)^2}$,
and would take the deviations as from other possible finite size effects.
However, if it is the case,
the significant deviations in $1.5\lesssim L\lesssim 3$ fm in the right figure may not be justified.
In the range of $L$ where
the lattice data in the left figure scale just as the solid line,
we should observe the same behavior in the right figure.
Because, in the case when quarks are heavy,
composite particles will be rather compact and we expect smaller finite size effects.

We can understand this behavior on the assumption
that this state is a resonance state rather than a scattering state.
In fact, while the data in the left figure rapidly go above the solid line as L is decreased,
which can be considered to arise due to the finite size of a resonance state,
the lattice data exhibit almost no volume dependence
in the right figure especially in $1.5\lesssim L\lesssim 3$ fm,
which can be regarded as the characteristic of resonance states.

\section{$(I,J^P)=(0,\frac12^+)$ channel}
\label{Positive}

\begin{figure}[htb]
\begin{center}
\includegraphics[scale=0.29]{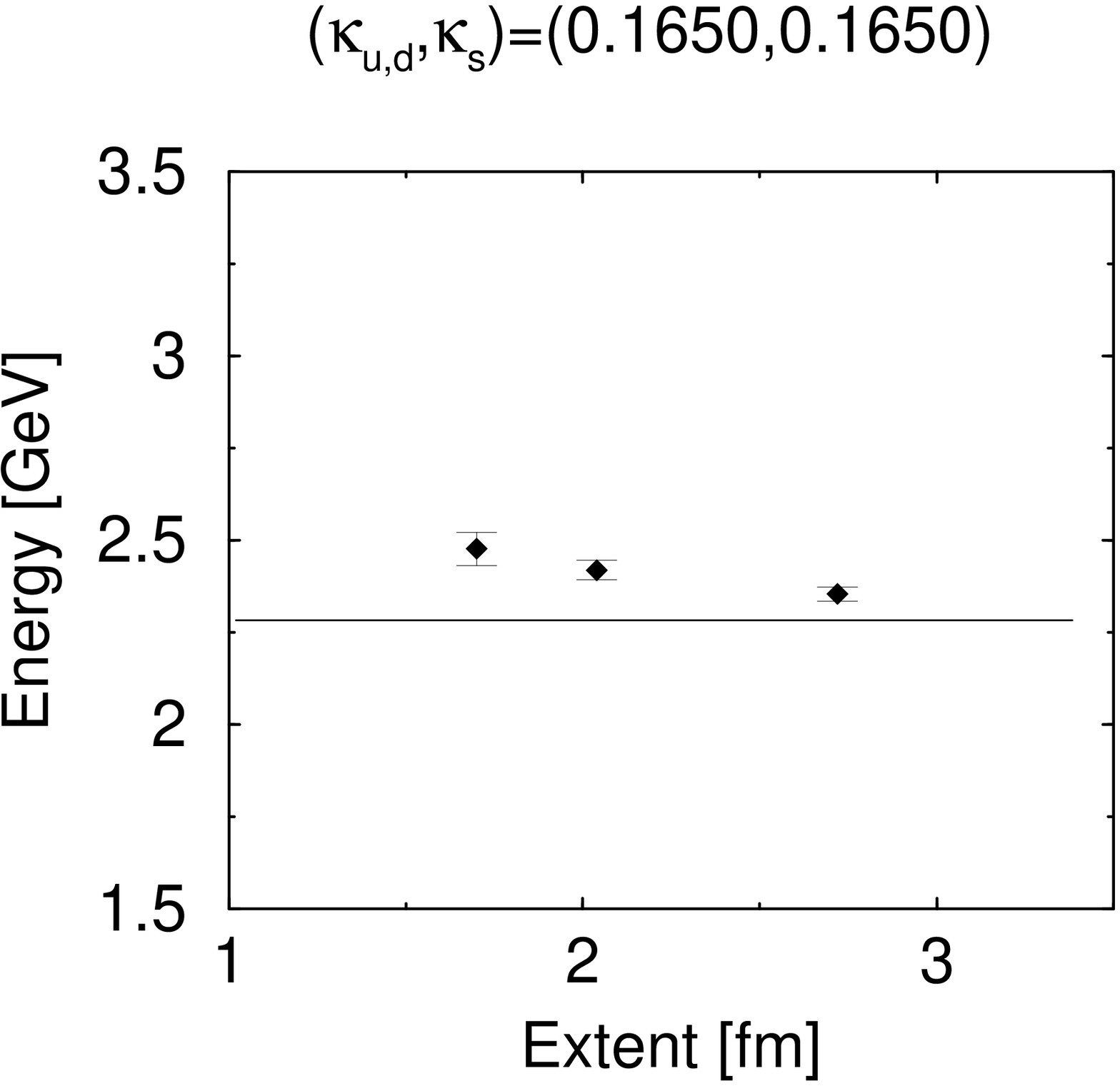}
\includegraphics[scale=0.29]{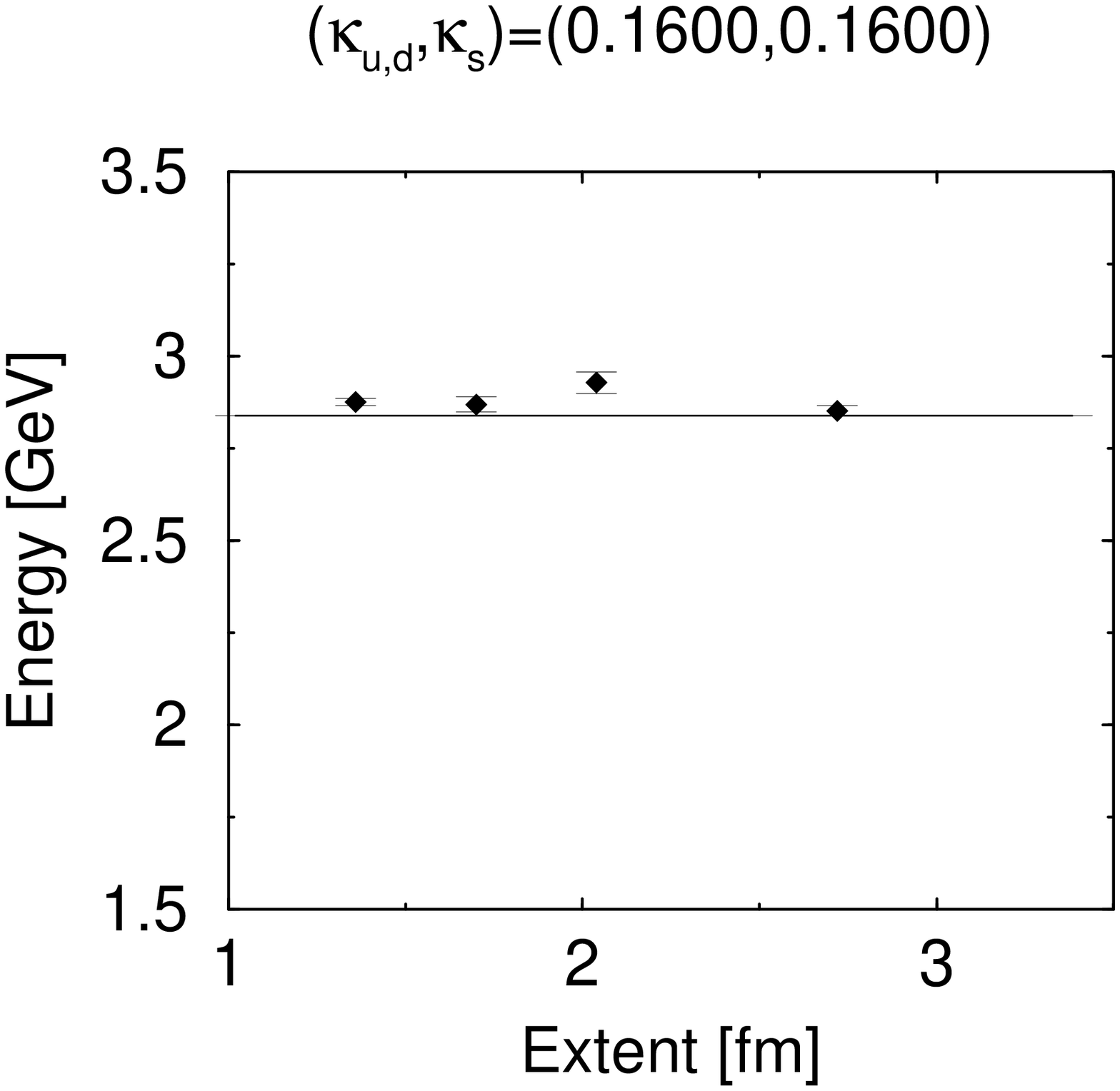}
\end{center}
\caption{\label{positiveSS}
The lattice QCD data in the $(I,J^P)=(0,\frac12^+)$ channel
are plotted against the lattice extent $L$.
The solid line denotes the simple sum $M_{N^*}+M_K$ of
the masses of the ground-state negative-parity nucleon $M_{N^*}$
and Kaon $M_K$.
}
\end{figure}
In the same way as $(I,J^P)=(0,\frac12^-)$ channel,
we have attempted to diagonalize the correlation matrix
on $(I,J^P)=(0,\frac12^+)$ channel
using the wall-sources $\overline{\Theta}_{\rm wall}(t)$.
In this channel, the diagonalization is rather unstable and we
find only one state except for tiny contributions of possible other states.
We plot the lattice data in Fig.~\ref{positiveSS}.
One finds that they have almost no volume dependence
and that they coincide with the solid line
which represents the simple sum 
$M_{N^*}+M_K$ of $M_{N^*}$ and $M_K$,
with $M_{N^*}$ the mass of the ground-state
of the {\it negative-parity} nucleon.
From this fact,
the state we observe is concluded to be the $N^*K$ scattering state
with the relative momentum $|{\bf p}|=0$.
It is rather strange
because the p-wave state of N and K with the relative
momentum $|{\bf p}|=2\pi/L$
will be lighter than
the $N^*K$ scattering state with the relative momentum $|{\bf p}|=0$.
We miss this lighter state in our analysis.
This failure would be due to the wall-like operator $\Theta_{\rm wall}(t)$:
The operator $\Theta_{\rm wall}(t)$
is constructed by the spatially spread quark fields 
$\sum_{\vec{\bf x}}q({\rm x})$ 
with zero momentum.
This may lead to the large overlaps
with the scattering state with zero relative momentum.
It is also desired to try another operator which couples
to p-wave NK scattering state.

\section{Summary}

We have performed the lattice QCD study of the S=+1 pentaquark baryons
on $8^3\times 24$, $10^3\times 24$, $12^3\times 24$
and $16^3\times 24$ lattices at $\beta$=5.7 at the quenched level
with the standard plaquette gauge action and Wilson quark action.
With the aim to separate states,
we have adopted two independent operators with $I=0$ and $J^P=\frac12$
so that we can construct a $2\times 2$ correlation matrix.
From the correlation matrix of the operators,
we have successfully obtained the energies of the ground-state and 
the 1st excited-state in $(I,J^P)=(0,\frac12^-)$ channel.
The volume dependence of the energies
shows that the 1st excited-state in this channel is 
rather different from a simple NK scattering state
and is likely to be
the resonance state located slightly above the NK threshold,
and also indicates that
the ground-state is the NK scattering state with the relative momentum
$|{\bf p}|=0$.
As for the $(I,J^P)=(0,\frac12^+)$ channel,
we have observed only one state in the present analysis,
which is likely to be a $N^*K$ scattering state
of the negative-parity nucleon $N^*$ and Kaon
with the relative momentum $|{\bf p}|=0$.

\end{document}